\documentclass[conference]{IEEEtran}

\usepackage[utf8]{inputenc} 
\usepackage[nocompress]{cite}   
\usepackage{microtype}
\usepackage{balance}

\usepackage{array}
\usepackage{multirow}
\usepackage{xcolor}
\usepackage{pgfplots}
\pgfplotsset{compat=1.16}
\usepackage{tikz}
\usetikzlibrary{patterns}
\definecolor{dark-red}{rgb}{0.3,0.1,0.1}
\definecolor{dark-green}{rgb}{0.1,0.3,0.1}
\definecolor{dark-blue}{rgb}{0.1,0.1,0.5}

\usepackage{url}
\usepackage[nameinlink]{cleveref}

\newcommand\copyrighttext{%
  \tiny \textcopyright 2020 IEEE. Personal use of this material is permitted. Permission from IEEE must be obtained for all other uses, in any current or future media, including reprinting/republishing this material for advertising or promotional purposes, creating new collective works, for resale or redistribution to servers or lists, or reuse of any copyrighted component of this work in other works. Cite this article as follows: R. Motschnig, M. Silber, and V. Švábenský, \textit{How Does a Student-Centered Course on Communication and Professional Skills Impact Students in the Long Run?}, in Proceedings of the 50th IEEE Frontiers in Education Conference (FIE '20). Uppsala, 2020. DOI: 10.1109/FIE44824.2020.9273962.}
\newcommand\copyrightnotice{%
\begin{tikzpicture}[remember picture,overlay]
\node[anchor=south,yshift=12pt] at (current page.south) {\fbox{\parbox{\dimexpr\textwidth-\fboxsep-\fboxrule\relax}{\copyrighttext}}};
\end{tikzpicture}%
}

\newcommand{\CSS}{\textsc{Communication and Soft Skills} }

\begin{document}

\title{How Does a Student-Centered Course on Communication and Professional Skills Impact~Students in the Long Run?
}

\author{\IEEEauthorblockN{Renate Motschnig}
\IEEEauthorblockA{\textit{Faculty of Computer Science}\\
\textit{University of Vienna} \\
Vienna, Austria \\
renate.motschnig@univie.ac.at}
\and
\IEEEauthorblockN{Michael Silber}
\IEEEauthorblockA{\textit{Faculty of Computer Science}\\
\textit{University of Vienna} \\
Vienna, Austria \\
michael.silber@univie.ac.at}
\and
\IEEEauthorblockN{Valdemar Švábenský}
\IEEEauthorblockA{\textit{Faculty of Informatics}\\
\textit{Masaryk University}\\
Brno, Czech Republic \\
svabensky@fi.muni.cz}
}

\maketitle
\copyrightnotice

\begin{abstract}
This Full Paper in the Research-To-Practice Category presents a long-term study about the effects of a student-centered course on communication and professional skills on students’ thoughts, attitudes, and behavior. The course is offered at a European university as part of a computer science master's program. This paper shares the design and challenges of a longitudinal study that reaches ten years behind and employs a mixed-methods approach. Besides presenting and interpreting the findings, we shed light on which features tend to stay on students’ minds and impact their way of being and acting in society. Moreover, we suggest implications for the design and practice in comparable courses to maximize constructive, sustainable effects, such as improved active listening, presentation skills, and openness to other perspectives. These are essential (not only) for computer science professionals. Our findings suggest that the course provided significant learning for the vast majority of respondents, including aspects such as presenting while keeping the other side in mind, managing one’s stress, and becoming less shy to speak up. All in all, we aim to contribute an evidence-based source of motivation for instructors in technically focused curricula who hold a student-centered stance.
\end{abstract}


\section{Introduction}
\label{sec:intro}

While teaching professional skills in computer science and other STEM (Science, Technology, Engineering, Mathematics) disciplines is often demanded for economic reasons, it is still more disputed than teaching exclusively technical subjects~\cite{Capretz2018}. To add evidence to the effectiveness and acceptance of a master-level course on professional skills, this paper sketches a course design and describes a mid- to long-term study on what graduates remember and take away from such a course. Thus, this work aims to complement related short-term studies such as~\cite{verdugo2013, Blume2009, Holzer2014} by researching sustainable effects, that is, what stays with graduates over the years that they apply in their jobs and possibly beyond.

From the perspective of curriculum design, the ACM/IEEE guidelines for computer science programs~\cite{Sahami2013computer} recommend promoting professional development. The authors argue that education must prepare students for the workplace in a ``holistic way'' and that ``soft skills (such as teamwork, verbal and written communication, time management, problem solving, and flexibility) and personal attributes (such as risk tolerance, collegiality, patience, work ethic, identification of opportunity, sense of social responsibility, and appreciation for diversity) play a critical role in the workplace''~\cite[p. 15]{Sahami2013computer}. They further state that ``Students should have opportunities to develop their interpersonal communication skills as part of their project experience'' (p. 24) and ``Graduates should have the ability to make effective presentations [\ldots]. They should be prepared to work effectively as members of teams. Graduates should be able to manage their own learning and development, including managing time, priorities, and progress'' (p. 25). These recommendations directly influenced the design of the course studied in this paper.

Other connections between information technology (IT) and professional skills were highlighted by Chou~\cite{chou2013fast}, who argued that a successful career substantially depends on soft skills. This holds true for IT professionals in particular but transcends to all fields. He argued that ``good verbal skills do not necessarily result in good communication skills. \textit{But people in the top tier do have one thing in common: they all excel in soft skills.}''~\cite[p. 21]{chou2013fast}. In the context of software engineering,~\cite[p. 1]{Capretz2018} observed and complained that professional -- or soft -- skills ``are often being overseen in professional work places and education.'' However, according to~\cite[p. 1]{Capretz2018}, soft skills are crucial in creating software and ``collaboration, communication, problem-solving and similar interpersonal and critical thinking skills that are expected from software engineering professionals.'' Like Zhang~\cite{Zhang2012}, we demand ``that computer science and software engineering curricula should put more emphasis on developing and assessing both hard and soft skills''.

The research most closely related to the current study concerns the students’ perceived short-term effects of a single course instance~\cite{Motschnig2019}. The major differences between the two studies are the mid- to long-term perspective in the current research, employing an online survey for data collection sent to almost 1,000 graduates, and utilizing a mixed-methods approach. The latter combined ordinal-scale questions with open questions. The detailed, qualitative responses were analyzed~\cite{mayring2014qualitative}. The findings allow us to draw conclusions on influential factors of the course design and pedagogical approach.

The target audience of this paper are instructors/facilitators, curriculum designers, educational researchers, and everybody interested in a student-centered way of promoting professional skills in higher education and STEM context. With these audiences in mind, the next section will sketch the researched  \CSS course.

\section{A Student-centered Way of Mediating Communication and Professional Skills}

\subsection{Pedagogical Basis}

The ``philosophy'' and design of the course are based on three major pillars summarized below:

\subsubsection{Experiential, student-centered learning}
It was practiced and researched by Carl Rogers and other researchers and educators such as~\cite{cornelius2009learner, froyd2008student, wright2011student, MOTSCHNIGPITRIK2013, hoidn2016student}. It was enhanced by technology~\cite{MOTSCHNIGPITRIK2013, bauer2006promotive, Derntl2004}, giving rise to Person-Centered technology-enhanced Learning (PCeL). Note that PCeL is a holistic approach to learning, addressing the whole person with their knowledge, skills, attitudes, and feelings. Therefore, it integrates a person’s cognitive and implicit functioning and so aims at sustainable learning that makes a difference to the person~\cite{rogersbecoming, motschnig2014person}.

\subsubsection{Dale’s cone of experience}
\cite[p. 105]{dale1969audiovisual} builds upon the theory that the more channels are activated by a stimulus, the bigger the retention. The basis of the cone, occupying the largest area, corresponds to learning with the highest retention rate. It is labeled as “Direct Purposeful Experience -- Go through the real experience.” In essence, the closer the learning stimulus resembles the real situation, the higher the retention or sustainability of the respective learning.

\subsubsection{ACM/IEEE guidelines for the design of computer science curricula}
\cite{Sahami2013computer} recommend a holistic approach and include professional skills and attitudes as described in~\Cref{sec:intro}.

\subsection{Course Goals and Course Description}
The primary goal of the course \CSS is to allow and motivate students to improve their competence in communication, teamwork, moderation, and other professional skills based on the students’ demands. In the course, students acquire knowledge and skills in active listening, person-centered communication, moderation techniques, interactive presentations, teamwork, and related skills in a holistic way, coupled to their job- and study-related needs and their solicited expectations.

The course with a group size of up to 20 participants is based on active PCeL in an open-error climate~\cite{MOTSCHNIGPITRIK2013, motschnig2019person}. The instructor acts as a facilitator and moderator who provides resources and moderates the initial block of the course on the topic of communication and community building, modeling the style that student teams would use for their upcoming moderation units. The focus lies on providing a constructive course climate based on genuineness, respect, and encompassing understanding~\cite{rogers1983freedom, motschnig2014person}, allowing students to feel safe and to open up about sharing their experiences.

Active inclusion of each participant is a priority, as is the shaping of active listening skills, evoking participants’ motivation for life-long development of their professional skills. Students work in teams of 3--4 people on a soft skills topic that they can choose after consultation with the whole group. Each team researches information on the topic and designs an interactive learning scenario in a self-directed way. This scenario is then consulted with the facilitator and subsequently moderated in a time span of 2.5 to 3 hours. Each moderation unit is followed by detailed oral, immediate feedback from the group, including the facilitator. Further feedback is submitted online in participants’ reaction sheets~\cite{MotschnigPitrik2014Patterns}.

The course is blocked into three units, each lasting one and a half days with a three-week break between the blocks. Assessment includes students’ active participation in face-to-face and online phases, the elaborated materials, the quality of the moderated unit, and the online self-evaluation. A detailed description of the course goals, tasks, the course mode, and assessment can be found at \url{https://cewebs.cs.univie.ac.at/ext-css/ws17/}.

\section{Research Questions and a~Mixed-Methods~Approach}
\label{sec:methods}

\subsection{Research Questions and Procedure}

This research focuses on the following questions:
\begin{itemize}
\item[RQ1] In how far students think that, in the long run, the course had helped them in communicating i) with their colleagues at work and ii) in their personal lives?
\item[RQ2] In how far students think that, in the long run, the course had helped them in connecting to people for professional and/or social benefit?
\item[RQ3] In how far students think that the course had motivated them to seek further development in the area of professional skills?
\item[RQ4] What is the students’ “anchor”, that is, the first memory of the course, and what memories had stayed on their minds in the long run?
\item[RQ5] Which valuable long-term effects of the course do students perceive in their i) working life and ii) personal life?
\item[RQ6] Based on students’ experience in the course, is there anything that influenced them (for example, impacted their thinking, attitudes, or behavior)?
\end{itemize}

To answer these questions, we designed an online questionnaire with open and closed questions. It was inspected by several students, revised, and emailed to all students who had attended the course \CSS in the last ten years (19 semesters, up to autumn 2008). These were a total of 938 students, and the time span for students who had attended the course in the most recent semester exceeded three months. These graduates were asked to participate in the online survey administered by the information system of the host university. About 230 students had opened the questionnaire, and 73 (7.8\%) filled it out and submitted it, 49 of them within the first 24 hours, the rest within about 3 weeks with one friendly reminder a few days before submission was closed.

\subsection{Participants and Their Demographic Data}

Before describing the analysis of the responses, we share facts about the student population. The vast majority of students attending the course studied Service Science Management and Engineering or Software Systems and Services Management. The age range in 2009 was between 20.7 and 25.3 years, with an average age of 22.8. There were 46 males and 4 females. In 2018, students were between 20.4 and 36.2 years old, with an average of 24.4 years. 216 males and 43 females were enrolled. Most students were Czech or Slovak, and 3--9 international students participated in the English-speaking group of the course each term. About two-thirds of the master's students were employed, a fair share of them on a full-time basis.

\subsection{Mixed Methods Data Collection}

To address the research questions, we used a mixed-methods approach: an online survey consisting of questions with predetermined, ordinal-scale grouped responses and questions with open, free-text responses. For the first three research questions (see~\Cref{fig:barchart}), we had offered response options: “not at all, slightly, moderately, much, very much, and not specified.” To illustrate the responses, we chose bar charts displaying the number of responses for each option.

While this quantitative part provides a gross estimate of the course’s effects at a glance, it does not allow to deduce information about students’ memories, gained competencies, particular impact factors, and, most importantly, the perceived effects of the course. This is why we had chosen an open response format in the survey for inquiries that would allow us to respond to research questions 4 to 6 (\Cref{tab1} and \Cref{tab2}). The questionnaire had been sent out bilingually in English and Czech to improve its comprehension for non-native English speakers. Responses were encouraged in English while also Czech and Slovak responses were accepted.

\subsection{Mixed Methods Data Analysis}

The data analysis focused on finding out which traces the course had left on students and whether they recalled anything that had a lasting effect on them. This is why we refrained from thematic analysis and instead chose the more detailed and fine-grained qualitative content analysis according to Mayring~\cite{mayring2014qualitative}. This method allowed us to mix inductive and deductive category formation. This proved helpful because, on the one hand, we had several categories on our mind when forming the research questions. On the other hand, we wanted to stay open to students’ responses and form subcategories from reading the responses. Also, the flexibility of Mayring’s method allowed us to apply shortcuts that saved us time and worked effectively, like paraphrasing responses while analyzing the data. Furthermore, we considered the frequencies of statements in each (sub-)category since these numbers mirror the weight of the respective impact on course graduates. This again allows us to draw conclusions about which content and instructional approach leaves long-term traces in students’ learning.

Mayring~\cite[p. 80]{mayring2014qualitative} proposes eight steps for performing qualitative analysis. Below, we describe our activities while following these eight steps with a few modifications that are in tune with the method that Mayring considers a process in which new decisions may impact the procedure.

\subsubsection*{Step 1: Research question, theoretical background}
While we had specified the research questions prior to any methodological decisions, step 1 made us more aware of the exploratory nature of this research: What is it in an academic, student-centered course that tends to sustain in students over the years? Also, the frequencies of related statements would illustrate information about what stayed with how many students.

\subsubsection*{Step 2: Establishment of a selection criterion, category definition, level of abstraction centers}
Due to finding a variety of relevant statements in the questionnaires, we chose to analyze responses to qualitative questions in all 73 cases. These amounted to a corpus of 8,873 words. 17 of the 73 respondents (23.3\%) responded in Czech/Slovak, and their responses were translated into English by the authors before the analysis. The unit of analysis was a meaningful phrase. This could be a part of a longer sentence or a few associated sentences. In sum, 745 units/meaningful phrases were distinguished and analyzed.

The category system was established inductively based on the research questions. Subcategories were formed following a mixed inductive-deductive process with some subcategories determined in advance and others derived from the responses. We deductively posed the categories: “anchor” or what spontaneously comes to mind; memories that stayed on graduates’ minds; valuable effects on working life and on personal life; the course’s impact on thinking, attitudes, or behavior; anything (for example, skills or knowledge) used as a consequence of attending the course; anything else worth sharing.

To increase reliability, two co-authors of this article engaged in the rating. One of them is the originator and instructor of the course, while the other is an advanced master’s student in computer science who had not attended the course and was expected to challenge the potentially hidden assumptions. Retrospectively, this indeed happened with a few statements. However, after careful consultation of each rating, the inter-rater agreement rose from the initial 75\% to 100\%. The most differences among the researchers concerned statements that fitted two (or more) subcategories. Since, for simplicity, we refrained from multiple categorization, we had to decide which subcategory would fit better. For example, one student remarked that the course was a total waste of time. This statement was categorized as “negative impression” due to the emotional wording. But the statement would equally fit the subcategory “course as such” since it concerned the whole course.

\subsubsection*{Steps 3--5: Working through the texts line by line, new category formulation or subsumption; Revision of categories and rules after 10-50\% of texts; Final working through the material}
In our process of establishing subcategories, each of the two researchers independently came up with a list of subcategories for each of the four sets of related survey questions. These proposals were discussed and combined into one list of candidates. Thereupon we sat together and went through about 10--15 responses to each of the questions and compared, discussed, and coded our categorizations in an MS Excel™ sheet. The remaining responses were categorized independently: one researcher selected a subcategory from the relevant subcategory-list, and the second researcher only saw the co-researcher’s choice once (s)he had entered their subcategory. Who rated first depended on the researchers’ momentary time resources and alternated. Subsequently, adapting one’s choice was possible, and all ratings that remained different were carefully discussed and turned into consensus. Hand in hand with assigning statements to subcategories, one researcher described the categories, and the other researcher checked them based on his/her understanding. Subsequently, both researchers marked those statements that appeared to be prototypical examples of their respective subcategory and associated the most filling ones with the respective subcategory.

\subsubsection*{Step 6: Building of main categories if useful}
For brevity, similar subcategories were grouped under an encompassing term. However, we did not cancel the original subcategories to preserve the information inherent in the original, detailed categorization. For example, the subcategories “active listening skills”, “presentation skills”, and “communication skills” were grouped under “all communication skills”. The subcategories “aspect of course as a whole” and “course as such” were grouped under “course-related” (compare \Cref{tab1} and \Cref{tab2}).

\subsubsection*{Step 7: Intra-/Inter-coder agreement check}
The intra- and inter-coder agreement and consistency checks happened iteratively throughout the process rather than just in one final step. During the final meeting, we resolved the remaining discrepancies on assigning the subcategories and agreed upon the format of presenting the content analysis findings.

\subsubsection*{Step 8: Final results, ev. frequencies, interpretation}
We agreed to consider frequencies of statements falling into the same (sub-)category since these would reflect the significance of the respective (sub-)category for the student population.

\section{Findings and Their Interpretation}
\label{sec:findings}

\subsection{Quantitative results (RQ1--RQ3)}

\begin{figure}[t]
\centering
\footnotesize
\begin{tikzpicture}
    \begin{axis}[
        width = 0.51\textwidth,
        height = 6cm,
        major x tick style = transparent,
        ybar=2*\pgflinewidth,
        bar width=8pt,
        ymajorgrids = true,
        ylabel = {Count},
        symbolic x coords={Not at all, Slightly, Moderately, Much, Very much},
        xtick = data,
        scaled y ticks = false,
        enlarge x limits=0.15,
        ymin=0,
        ymax=25,
        nodes near coords,
        every node near coord/.append style={color=black},
        legend style={
                at={(-0.107,1.825)},
                anchor=north west,
                column sep=1ex,
                cells={align=left}
        }
    ]
        \addplot[style={gray!10,fill=gray!10,mark=none},postaction={pattern=north east lines}]
            coordinates {(Not at all, 4) (Slightly, 10) (Moderately, 21) (Much, 21) (Very much, 13)};
        \addplot[style={gray!40,fill=gray!40,mark=none},postaction={pattern=crosshatch}]
            coordinates {(Not at all, 5) (Slightly, 14) (Moderately, 21) (Much, 22) (Very much, 8)};
        \addplot[style={gray!70,fill=gray!70,mark=none},postaction={pattern=dots}]
            coordinates {(Not at all, 15) (Slightly, 13) (Moderately, 17) (Much, 12) (Very much, 12)};
        \addplot[style={gray!95,fill=gray!95,mark=none},postaction={pattern=horizontal lines}]
            coordinates {(Not at all, 6) (Slightly, 13) (Moderately, 14) (Much, 22) (Very much, 15)};
        \legend{
        {RQ1i: Do you think that, in the long-term perspective, the course\hspace*{0.75mm}\\helped you in communicating \textit{with your colleagues at work}?},
        {RQ1ii: Do you think that, in the long-term perspective, the course\\helped you in communicating \textit{in your personal life}?},
        {RQ2: Do you think that, in the long-term perspective, the course\hspace*{1.5mm}\\helped you \textit{to gain useful contacts} (for example, for studying\\for finals, finding a job, or making friends)?},
        {RQ3: Do you think that the course contributed to your further ac-\hspace*{0.1mm}\\tive interest in the area of soft skills (for example, reading\\books or taking other courses)?}}
    \end{axis}
\end{tikzpicture}
\caption{Descriptive statistics of the effects of the course as indicated by participants who responsed to RQ1i, RQ1ii, RQ2 and RQ3 ($n$ = 73).}
\label{fig:barchart}
\vspace{-5mm}
\end{figure}
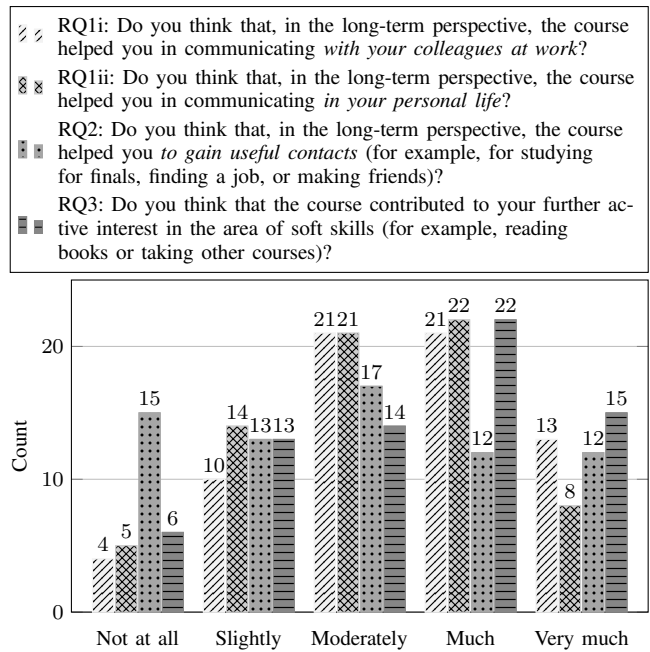

To provide a gross estimate and overview of the mid- to long-term effects of the course, the students were asked four questions depicted above~\Cref{fig:barchart}. Only questions RQ1i about improving professional communication and RQ3 concerning motivation to continue developing professional skills addressed essential explicit course objectives. RQ1ii about improving communication in personal life was added. We assumed it to be a positive side effect of the course that would illustrate the pervasiveness of constructive communication as an expression of attitude rather than just a technique. RQ2 on “useful contacts” was included to provide us with data on community-building and networking in or via the course~\cite{bandura1977social} that students frequently had noticed as a unique feature of the course.

\definecolor{bblue}{HTML}{4F81BD}
\definecolor{rred}{HTML}{C0504D}
\definecolor{ggreen}{HTML}{7ABB48}
\definecolor{ppurple}{HTML}{9F4C7C}

\Cref{fig:barchart} shows that students received the course objectives (RQ1i and RQ3) somewhat more than the positive-side effects (RQ1ii and RQ2). Regarding professional communication, 34 out of 69 (49.3\%) of the respondents indicated that the course had helped them “much” or “very much” to improve their communication at work. In contrast, only 4 (5.8\%) did not perceive any gain in their work-related communication that they would account to the course. A similar but not as distinct result accrues for personal communication, where 30 out of 70 respondents said that the course had helped them “much” or “very much” to improve their communication in personal life. Only 5 (7.1\%) did not perceive any course-related gain in their private communication. Turning to establishing contacts, 24 out of 69 respondents felt that the course had helped them in this respect, while 15 (21.7\%) did not report any course-induced support on networking. While this is the relatively weakest aspect inquired, it may depend on the graduates’ personalities in how far they value contacts but also on their situation (staying at the university or leaving, often also to a different country).

We were most happy to see that the course motivated 37 out of 70 (that is, 52.9\%) of the respondents “much” or “very much” to continue developing their professional skills, while just 6 (8.6\%) did not feel inspired by the course to proceed in their professional development. Note that this can have several reasons reaching from a lack of interest to a strong interest already in place before attending the course.

Below, we discuss the qualitative outcomes of the content analysis in terms of (sub-)categories, representative statements, and the frequency of statements in a (sub-)category. The aim is to illustrate the course’s mid- to long-term impact on graduates, along with insights we gained from this research.

\subsection{Graduates’ memories (RQ4)}

The categories “what spontaneously comes to graduates’ minds” (RQ4i) and “memories of the course” (RQ4ii) deal with the “anchor” or first memory that enters the respondents’ minds when thinking of the course and the further elaboration of their memories, respectively. \Cref{tab1} shows that the further elaboration of memories received about twice as many responses (187) than the first memory (95). Interestingly, the most frequent statements in both categories addressed specific aspects of the course or the course as such (33 as first memories and 57 as elaborated memories). For example, course graduates remembered the intensive teamwork and the immediate practical application of “individual soft skills learned in the course”, the “great atmosphere”, “sitting in a circle”, or “the non-traditional way of lectures”.

The second most frequent subcategory were spontaneous (19) and elaborated (35) memories about “activities”. Examples concern the close collaboration in one’s team, moderating a topic while keeping the audience engaged, preparing materials for schoolmates, “talking in front of a lot of people”, and getting feedback. Group activities were remembered far more frequently than other activities, corroborating theories about the importance of relationships and social interaction~\cite{Zhang2012, rogers1983freedom}.

Memories about mediated “competences or skills” (the two terms are used interchangeably) were the third most frequent (24) among elaborated memories. Examples of statements on remembering competences/skills are “basics of active listening”, ways to “communicate with people, self-confidence, or non-verbal communication”. Among spontaneous memories, the subcategory “positive impression” (14) ranked third. Ad hoc memories of positive impressions concerned “the great time spent together”, “fun”, or “the awesome atmosphere”, mirroring the prevalent constructive climate.

\begin{table}[t]
\renewcommand{\arraystretch}{1.25}
\setlength{\tabcolsep}{1mm} 
\caption{Graduates’ Spontaneous (RQ4i) and Elaborated (RQ4ii) Memories of the Course}
\begin{center}
\begin{tabular}{|p{2.1cm}|p{4.5cm}|r|r|}
\hline
\multicolumn{4}{|l|}{RQ4i: Spontaneously, what is the first thing that comes to your mind?} \\
\multicolumn{4}{|l|}{RQ4ii: Can you elaborate on what you remember from the course?} \\ \hline
\textbf{Subcategory} & \multirow{2}{*}{Prototypical example} & \multicolumn{2}{l|}{Stmnt. count} \\ \cline{3-4}
\textbullet\ sub-subcategory & & RQ4i & RQ4ii \\ \hline\hline

\textbf{Course-related} & & \textbf{33} & \textbf{57} \\ \hline
\textbullet\ aspect of course as a whole  & “I remember that we have worked more in teams compared to other courses and in the process applied individual soft skills learned in the course.” & 28 & 47 \\ \hline 
\textbullet\ course as such  & “It was a good guide to soft skills everybody needs in his or her life.” & 5 & 10 \\ \hline\hline

\textbf{Activities} & & \textbf{19} & \textbf{35} \\ \hline
\textbullet\ group activity & “Working closely with other members of my presentation team.” & 13 & 16 \\ \hline 
\textbullet\ presentation & “Teaching a topic to a group of people while keeping them engaged.” & 2 & 11 \\ \hline 
\textbullet\ activity & “The most interesting thing was to prepare materials for my schoolmates regarding [a] chosen topic. It's quite an experience for my life.” & 4 & 8 \\ \hline\hline

\textbf{Competences} & & \textbf{7} & \textbf{24} \\ \hline
\textbullet\ other competence / skill & “[The course is] not about knowledge, but rather, it gave me a way to communicate with people I see for the first time. It gave me self-confidence and showed that communication can be effective in other ways than verbally. In fact, it gave me more than most of the other courses of the Master Study.” & 3 & 15 \\ \hline 
\textbullet\ active listening & “I still recall the basics of active listening and its importance. During my work I often come back to this topic and search for additional material.” & 4 & 9 \\ \hline\hline

\textbf{People-aspect} & “I remember that with the other participants we became quickly a great group where everybody supported each other.” & \textbf{12} & \textbf{20} \\ \hline\hline

\textbf{Knowledge item} & “Also the topic of conflict resolution was helpful for me later.” & \textbf{4} & \textbf{20} \\ \hline\hline

\textbf{Positive impression} & “Positive feelings about cooperation in my team and with teachers. I am still meeting my team members from time to time.” & \textbf{14} & \textbf{17} \\ \hline\hline

\textbf{Negative impression} & “Overall, the course was disproportionally long, lacking practical exercises such as conflict situations, presentations, giving and receiving feedback.” & \textbf{1} & \textbf{6} \\ \hline\hline

\textbf{Other} & “During my work I do often come back to this topic and search for additional books and material.” & \textbf{5} & \textbf{8} \\ \hline\hline

Statement sum & & 95 & 187 \\ \hline
No response & & 0 & 3 \\ \hline
\end{tabular}
\vspace{-5mm}
\label{tab1}
\end{center}
\end{table}

Next in terms of statement frequencies came memories of the “people-aspect”, specified as “Anything connected to the human players (students, lecturers) of the course”. People-related aspects often concerned peers, such as in “with the other participants, we became very quickly a great group where everybody supported each other”. Occasionally, people-aspects referred to teachers, such as in “I really appreciated that the teachers weren't teaching but guiding and moderating the discussions”. This nicely reflects the dynamics of student-centered education in which teachers act as facilitators, and students are at the forefront most of the time.

With 20 statements, memories of “knowledge-items” – specified as “any particular content or reference or concept that students remember” – were far more frequent in the elaborated memories category. Sample statements were: “Carl Rogers”, “the basics of the Person-Centered Approach”, “Pomodoro time-management”, or “the importance of the power gestures”.

From the other subcategories in \Cref{tab1}, let us examine “negative impression”, specified as “Anything that left a negative feeling, attitude, or trace in relationship to the course“. For one student, “waste of time” was the “anchor” to the course. Elaborated memories of negative impressions (6) were, for example: “The course was disproportionally long.” and “My group for the assignment was horrible [\ldots], they didn't care about the outcome of the assignment and in the end we had long and boring theoretical part and a game that was not working and it kept people from going home on time.” This illustrates the range of power dynamics in student teams, requiring thoughtful solutions. Students had been offered consultations throughout the course, but not everyone took this opportunity.

Summarizing, the graduates’ memories mirror the characteristics of student-centered learning. The distinguishing features of the course, such as students’ active roles, the constructive climate, and emphasis on group activities are remembered more frequently than pure knowledge items, even though the latter remain present in several graduates’ minds. Positive impressions (31 statements) by far outweigh negative ones (7), which mainly concerned the length of the course blocks, peer-teaching, and lack of rigor in the grading criteria. Since the course was compulsory, a few students apparently did not manage to open up to its unorthodox style. Reflection groups among instructors might help to better understand and deal with students’ blocked motivation.

\subsection{The effect on professional and personal lives (RQ5--RQ6)}

We now analyze graduates’ responses to three related research questions. (RQ5i): “Did you take something from the course that was later valuable for you in your working life?” (RQ5ii): “Did you take something from the course that was later valuable for you in your personal life?” and (RQ6): “Based on your experience in the course, is there anything that influenced you (for example, it had an impact on your thinking, attitudes, or behavior)? If so, please describe it.” These questions gave rise to the three categories: the takeaway for working life, the takeaway for personal life, and the influence on the person. The respective subcategories are listed in \Cref{tab2} and described below. In accordance with the course goals, we focus on the courses’ effects on the working life of course-graduates.

\begin{table}[!th]
\renewcommand{\arraystretch}{1.25}
\setlength{\tabcolsep}{1mm} 
\caption{Graduates’ Takeaways for Working Life (RQ5i), for Private Life (RQ5ii), and Influences on Person (RQ6)}
\begin{center}
\begin{tabular}{|p{2.1cm}|p{3.9cm}|r|r|r|}
\hline
\multicolumn{5}{|l|}{RQ5i: Did you take something from the course that was later valuable} \\[-1mm]
\multicolumn{5}{|l|}{for you in your \textit{working life}?} \\
\multicolumn{5}{|l|}{RQ5ii: [\ldots] that was later valuable for you in your \textit{personal life}?} \\
\multicolumn{5}{|l|}{RQ6: [\ldots] that influenced you (for example, it had an impact on your} \\[-1mm]
\multicolumn{5}{|l|}{thinking, attitudes, or behavior)? If so, please describe it.} \\ \hline
\textbf{Subcategory} & \multirow{2}{*}{Prototypical example} & \multicolumn{3}{l|}{Statement count} \\ \cline{3-5}
\textbullet\ sub-subcategory & & RQ5i & RQ5ii & RQ6 \\ \hline\hline

\textbf{All comm. skills} & & \textbf{45} & \textbf{40} & \textbf{13} \\ \hline
\textbullet\ communication skills  & “An ability to communicate the tasks more clearly while having the other receiving side in mind.” & 18 & 14 & 7 \\ \hline 
\textbullet\ active listening skills & “Being just open to other people and letting themselves to figure out what is their problem is powerful. I try to listen and ask more at my work and also at home.” & 14 & 23 & 6 \\ \hline
\textbullet\ presentation skills & “I learned how to present in front of people and deal with stress.” & 13 & 3 & 0 \\ \hline\hline

\textbf{Other skills} & “I use some of the time management/"Getting things done" rules in my working life.” & \textbf{17} & \textbf{8} & \textbf{4} \\ \hline\hline

\textbf{Specific effect, insight, and learning} & “I also breathe deeper when I dislike an opinion of a co-worker very much in order to calm myself so it is easier to dismiss the matter and not get into an argument.” & \textbf{17} & \textbf{11} & \textbf{14} \\ \hline\hline

\textbf{People-related aspects} & “Yes, in my opinion, it is important to maintain warm relationships with colleagues.” & \textbf{14} & \textbf{9} & \textbf{14} \\ \hline\hline

\textbf{Specific theory / knowledge} & “Team roles by Belbin” & \textbf{9} & \textbf{4} & \textbf{3} \\ \hline\hline

\textbf{Perceived unspecific impact of course} & “Nothing in particular comes to mind, but it was a nice practice -- as a team leader I use communication skills every day.” & \textbf{7} & \textbf{9} & \textbf{12} \\ \hline\hline

\textbf{Self-competences} & “It taught me self-confidence in preparing presentations and expressions.” & \textbf{6} & \textbf{9} & \textbf{5} \\ \hline\hline

\textbf{Course-related} & “The course was not purely about acquiring new skills, but also about making the existing ones sharper.” & \textbf{5} & \textbf{3} & \textbf{7} \\ \hline\hline

\textbf{Specific activity} & “Practicing the visit at company boss wanting pay raise was a valuable experience, getting great feedback and tips. Trying the teamwork as a CERT unit during cyber-attack was very exciting as well.” & \textbf{4} & \textbf{1} & \textbf{0} \\ \hline\hline

\textbf{Relationships with others} & “First listen and then react non-judgmentally/neutrally.” & \textbf{4} & \textbf{7} & \textbf{8} \\ \hline\hline

\textbf{Other} & “No.” & \textbf{3} & \textbf{8} & \textbf{8} \\ \hline\hline

Statement sum & & 131 & 109 & 88 \\ \hline
No response & & 4 & 6 & 10 \\ \hline
\end{tabular}
\vspace{-5mm}
\label{tab2}
\end{center}
\end{table}

By far, most graduates indicated a valuable takeaway on their communication skills at work (45), and almost as many statements (40) described a gain for communicating in their personal lives! Interestingly, “active listening skills” in personal life with 23 statements even outnumbered the “active listening skills” sub-subcategory in working life (14). Example statements regarding communication skills at work concerned becoming “more comfortable communicating with people at work”. The course particularly focused on active listening, and this is nicely reflected in typical statements. For example, a respondent wrote: “Active listening and non-verbal communication (not only) reading skills -- I have implemented them into my interactions and used them since the course and I believe they make me a better colleague/co-worker.” Another graduate responded: “I realized how important listening is [\ldots] and I use it all the time to show to other people that I care.” In particular, the last quote illustrates that communication skills are permeable, and their impact crosses professional borders. Regarding presentation skills, statements addressed, for example, “how to present things but still keep the focus of the audience” or becoming “less stressed when presenting”.

Besides communication, the subcategory “other skills” – specified as ``any (aspect of a) competence/skill that students report to have acquired or deepened as a result of the course'', received a considerable number of statements (17) in the “working life” category. These addressed skills/competencies such as leadership, time-management, assertiveness, conflict resolution, and teamwork. Sample statements particularly relevant for computer science graduates were: “I used more assertiveness as a tester (e.g., when I needed to tell developers about a bug I found)” or learning “to deal with stress”.

The subcategory “Specific effect, insight, and learning” – defined as “any takeaway perceived by graduates as a particular effect, insight, and learning in connection with the course” – had the same frequency (17) in the work context as “other skills” and even a higher frequency in the personal context (11) and influence on the person (14). Work-life examples were: “read body language of others and recognize my own body language” or “I'm a technical and semi-introverted type so I had to go out of my comfort zone, and I’m glad I did.” A statement illustrating the influence on the person was: “Since then, I know when my leads want to avoid answering my question or use the tactics to manipulate me.” These examples show that different students took away different learnings, underlining the large repertoire students had in the course and used for sustainable skills as well as the situations they encountered.

The subcategory of “People-related aspects” attracted 14 statements in both the work and personal category. As an example of the former, consider ”Thinking about co-workers' perspective about things” as an essential work-related capacity. Influence on the person is exemplified by the statement: “I think people from the course inspired me. I met a lot of interesting people who had a different way of thinking and organizing their lives. I pushed myself forward when I saw what those people were doing or willing to do.” Probably, similar experiences were more strongly symbolized in graduates’ minds than takeaways of the “specific knowledge/theory” subcategory, specified as “any takeaway attributed to a particular knowledge item in connection with the course,” and capturing 9 statements in the work category. Almost as many (8) items were assigned to the “relationship with others” subcategory regarding the influence on the person. There, statements pointing to an increase of openness and respect towards other perspectives were characteristic, such as: “It is always great to meet new people and learn how they think and what opinions they have. It always gives you another perspective.” \Cref{tab2} lists the remaining subcategories and typical statements.

In summary, the considerable numbers of reported takeaways for “working life” (131) and also “personal life” (109) signify that the course left a sustained trace on graduates in terms of subcategories listed in \Cref{tab2}. The statement frequencies further indicate that skills/competencies by far outnumbered the pure knowledge items, even though these were present as well, along with several specific learnings and a sense of benefiting from the course without specifying it. Furthermore, several takeaways like communicating tasks more clearly, keeping the receiving side in mind, managing stress, or knowledge about team roles appear particularly useful for working in IT.

Aside from the skills, the category “influence on the person” captured 88 statements, meaning that the course even left a trace on graduates’ thinking, attitudes, or behavior. Interestingly, in this category, people-related aspects and specific insights were most frequently mentioned, confirming the high value of social influence and freedom to learn on personal development.

The coverage of knowledge, skills, \textit{and} attitudes is congenial with student-centered, experiential learning according to~\cite{rogersbecoming, cornelius2009learner}, as well as with the specified goals of the \CSS course.

\subsection{Perceived sustainable effects and specifics}

To find out specific skills or insights students used based on the course, we asked the question: “Is there something from the course (for example, skills or knowledge) that you use as a consequence of attending the course?”. In sum, 75 statements were identified, out of which only 4 respondents answered “no” or similar. Unsurprisingly, “active listening” took precedence in terms of frequencies. Overall, the responses in the category “Something used as a consequence of attending the course” are consistent with the takeaways discussed above, yet more concrete and thus reflect the concrete effects of the course.

Finally, we wanted to give respondents the opportunity to mention whatever they wanted. This gave rise to the category “Anything else to share” that attracted 60 statements. The majority (17) was associated with the subcategory “Course as such” and emphasized the special nature, people, or atmosphere of the course. The subcategory “Gratitude” captured 11 statements, tightly followed by “Other” (9), “Positive impression” (8), and “Feedback” (6). At the end, the subcategory “Negative impression” came to hold the single statement: “The course was totally unluckily formed”, expressing the respondent’s discomfort with the course experience.

\section{Discussion}

\subsection{Limitations}

A shortcoming of our study is mainly the questionnaire response rate of 7.8\%. However, the rate cannot be calculated precisely since we do not know how many graduates were reached by the emails with the survey link. Unfortunately, some international students were not reached due to expired email accounts. With extra effort, their current email addresses could be found to contact them in a second, follow-up wave. In any case, we used all responses for the content analysis, capturing each statement by each of the 73 respondents.

Another shortcoming is that (due to space limitations) we did not distinguish between responses from early graduates (from 2009) and the recent ones. Since memories decline over time if not refreshed, differentiating between early and late graduates might reveal further information. The same holds for distinguishing between the groups conducted in English and holding international students versus the purely Czech/Slovak groups.

We also shortened Mayring’s procedure~\cite{mayring2014qualitative} of performing qualitative content analysis by skipping the step of paraphrasing each statement before its categorization. Instead, we used prototypical statements as an orientation that helped us to minimize the time spent on the analysis.

\subsection{Contribution}

Despite the limitations above, the mixed-methods approach allowed us to respond to all research questions listed in~\Cref{sec:methods}. The first three of them were researched quantitatively. It became evident that the explicit course goals, namely improved communication skills in the work context and motivation for life-long learning of professional skills outweighed other aspects such as establishing useful contacts and improving communication in personal life. The concordance of the results with the course goals confirms that formulating goals is important to establish focus throughout a course.

The results also show that for the vast majority of participants, the main course goals were achieved \textit{in a sustained form} at least to some degree. We attribute this primarily to the student-centered, holistic approach taken in the course. Students were highly engaged, could choose soft-skill topics they found relevant, prepared moderation units in a small, self-organized team, and actually worked with the group most of the time~\cite{Zhang2012}. This mode resonates well with the basis of Dale’s cone of experience: students went through the experience and received honest feedback from the group immediately.

While the findings regarding all research questions were discussed in~\Cref{sec:findings}, let us compare how the findings of the current study compare to the short-term effects of a single course instance reported in~\cite{Motschnig2019}. Interestingly, we observe a vast consistency of several effects of the course. In the short term, students had indicated that they were going to apply the learning in real life. Indeed, the numerous takeaways for “working- and personal life” (in sum 240!) nicely testify this tendency. Moreover, the short-term finding of an identified “positive change” in students was sustained, as can be seen from the versatile yet coherent category of “influence on the person” with the remarkable statement frequency of 88. Last but not least, students’ short-term observation that they had become better active listeners endured in several graduates. It reflected itself in high frequencies of the “active listening” subcategory in various categories such as memories, takeaways for personal and working life, and influence on the person.

While more research is needed, the resonating consistency of essential effects of the course between the short- and mid- to long term perspective allows us to draw the following conclusion. In student-centered, experiential courses grounded in a value-base of openness, honesty, respect, and deep understanding, the learning and personal growth of students are strengthened or initialized in the intensive course experience. Furthermore, it carried on for years, for the vast majority of participants who manage to open-up to the experience~\cite{rogers1983freedom}.

\subsection{Impact}

This contribution impacts higher education practice and research in the field of professional skills, student-centered courses, and mixed-methods research. Essentially, the researched course had started as an optional course with one group in 2006. Students grasped its relevance for the field and voted for it to become mandatory in their Service Science Management and Engineering curriculum. Gradually, the course became a sustainable, student-centered course with about seven groups per year.

The multiplying effect is achieved by offering the most gifted students to tutor the course with more experienced facilitators and, subsequently, to lead their own groups along with getting collegial supervision. This aligns with the Teaching Lab initiative at the Faculty, whose goal is to train new teaching assistants~\cite{ukrop2020}. Moreover, the generic course-design is “stable” as primary  course objectives tend to be reached for most students, and drop-out is minimized.

A justified question is how far researching the effects of the computer science master's course \CSS is relevant to computer science education. In fact, the course concept and goals are transferrable to any field. What matters centrally, however, is the students’ and facilitators’ orientation in the target discipline, along with the student-centered philosophy that allows students to choose the major course themes according to the felt need or interest to improve. Most often, this happens in the context of their (desired) job or a study project. In both cases, the real interest or demand arising from students’ jobs stands in the foreground and attracts the majority’s context – in our case, \textit{computer science}. Visiting students from other schools, nations, or cultures add the opportunity of getting an interdisciplinary or intercultural perspective. Thus, in a nutshell, it is the people who tune the course experience in the direction of a particular discipline.

Our research confirmed several graduates’ takeaways along the qualifications demanded by the guidelines for computer science curricula~\cite{froyd2008student}. Empirical research~\cite{Groeneveld2020} also shows how important these qualifications are. Consequently, we conjecture that while several aspects of our research reach beyond a particular discipline, this study also contributes to the special case of computer science education. For sure, it formed a firm basis for the facilitators’ own learning, (inter-)personal growth, and future development of the course.

\section{Conclusion and Further Work}

This work investigated students’ mid- to long-term takeaways from the master-level course \CSS via a mixed-methods approach. The quantitative results and qualitative findings were highly consistent and complemented each other. While the former provided an overview of how intensively essential course goals have been met, the latter illustrated takeaways and memories from the graduates’ perspectives using their own statements. We categorized them for conceptual coherence and counted them to estimate the “weight” of categories and subcategories.

In a nutshell, with a few exceptions, the course provided significant, long-lasting learning of skills, attitudes, and knowledge. These start from better active listening skills and reach to increased openness to the perspectives of others and valuing diversity, group activities, and creative, self-organized learning on one’s own and with peers.

Further work will organize a second wave of the online survey, reach international students who may have missed the original emails, and analyze responses based on gender and the year of course attendance. Also, the course design will be adapted to include video-conferencing experience that gains importance in the era of digitalization and occasional necessary social distancing, as for example, in the times of pandemics.

\section*{Acknowledgment}

In part, this research was supported by ERDF ``CyberSecurity, CyberCrime and Critical Information Infrastructures Center of Excellence'' (No. CZ.02.1.01/0.0/0.0/16\_019/0000822) and the project ``Soft Skills for IT-People'' by the Zero Outage Industry Standard (\url{www.zero-outage.com}). We appreciate this support and sincerely thank our students for engaging in the course and survey. We also thank the co-facilitators of the course, especially Vlasta Šťavová and Petra Kalábová who helped to design the early drafts of the questionnaire. 

\balance
\bibliographystyle{IEEEtran}
\bibliography{references}

\end{document}